\documentclass[12pt]{article} 

\usepackage{amsmath} 
\title{Is a glowing LED useful for accuratare determination of Plank's constant?}
\author{Chetan Kotabage\\
Department of Physics\\
KLS Gogte Institute of Technology\\
Belgavi 590008, India\\
Email:
cvkotabage@git.edu}
\begin{document}
\maketitle

\begin{center}
Abstract
\end{center}
 
Plank's constant is a fundamental constant in Physical sciences and Millikan received Noble prize for obtaining it experimentally in 1923. As the technology of solid state electronic devices developed, an experiment was designed and adapted to determine the Plank's constant using LED \cite{1}, \cite{2}.  Herrmann and Sch$\ddot{\text{a}}$tzle \cite{3} questioned the validity of this experiment on the basis of the diode equation and their approach was further discussed by Morehouse \cite{4}. In this article, the validity of this experiment is revisited.

\section{Introduction}
The relation between energy of a photon and kinetic energy of electron, which has been ejected by the photon in the photoelectric effect, has been instrumental in experimental determination of Plank's constant. Experimentally, multiplication of charge of electron and slope of stopping potential vs. frequency of radiation line gives Plank's constant \cite{5}. A somewhat reverse process, which takes place in LED, has been the basis of the experimental determination of Plank's constant using LED.

\section{Determination of Plank's constant using LED}
 In LED, a photon is emitted when an electron jumps from conduction band to valence band to recombine with a hole. The minimum energy of a photon emitted in this process is equal to the band gap energy, $E_g$. If the wavelength, $\lambda$, of this photon is known precisely, then a measurement of band gap energy can give the Plank's constant as
\begin{equation}
h=\frac{\lambda E_g}{c}\label{band gap energy}\;.
\end{equation}
The use of LED to find Plank's constant is based on this equation. In the experiment, I-V characteristics of LED are measured. The intercept of tangent of I-V curve on the voltage axis, which called as forward voltage $V_F$, is assumed to be equal to $E_g/q$, where $q$ is the charge of electron.   
 Thus, from eq.\eqref{band gap energy}, the Plank's constant is given as
\begin{equation}
h=\frac{\lambda V_F q}{c} \label{h by diode}\;.
\end{equation}

\section{Drawbacks of the method}

The experimental determination of Plank's constant by above method has the following drawbacks.

\subsection{Wavelength of light}

For accurate calculation of Plank's constant using eq. \eqref{h by diode},  the longest wavelength of light should be considered\footnote{ For some LED's, transition takes place when the electron is occupying a level below the lowest energy level of conduction band. So, such cases are not considered here.}. The wavelength of light depends on energy difference between a level of conduction band and a level of valence band,  which are involved in the transition of electron from conduction band to valence band. The band gap energy, $E_g$, has lowest energy difference because of energy difference between the lowest level of conduction band and the highest level of valence band. Thus, apart from photons of energy $E_g$, all other photons emitted by LED will be of higher energy than $E_g$. Since the wavelength of light is inversely proportional to energy of corresponding photon
\begin{equation}
E=\frac{hc}{\lambda}\;,
\end{equation}
a photon of energy $E_g$ will correspond to the longest wavelength of light.  Since eq. \eqref{h by diode} is obtained for photon of energy $E_g$, the longest wavelength of light must be considered. 

Usually, the calculations involve the wavelength specified by the manufacturer, which is not essentially the longest wavelength emitted by LED. Thus, lower values of wavelength of light in eq. \eqref{h by diode} results in underestimation of the Plank's constant. 

\subsection{Forward voltage}

Forward voltage is determined by the intercept of linear part of I-V characteristics on the voltage axis. This technique of defining forward voltage is incorrect as the intercept is much less than the ratio $E_g/q$. At forward voltage, diffusion of electrons and holes across the junction begins due to decrease in junction potential. At this voltage, respective conduction and valence bands of n and p type semiconductors shift closer by lower magnitude. Hence, the product of charge of electron and forward voltage of a diode is much less than band gap energy.

As mentioned in \cite{6}, the product of limiting forward voltage of a diode and charge of electron is slightly less than the band gap energy because of the highest possible shift between respective conduction and valence bands of n and p type semiconductors.  Hence, instead of forward voltage, limiting forward voltage is a more appropriate physical quantity to estimate Plank's constant.
\section{Conclusions}
Though use of LED to determine Plank's constant gives values that are close to Plank's constant, the experimental procedure is erroneous.  A more accurate procedure is to measure longest wavelength of light emitted by LED and utilize it along with limiting forward voltage to calculate Plank's constant. 

Since limiting forward voltage is marginally lower than band gap energy, such correction will not give accurate Plank's constant but rather refine the result. Hence, to determine Plank's constant accurately, band gap energy should be determined precisely.  

\section{Acknowledgment}
I thank anonymous referee of Resonance-Journal of Science Education. 


%
%
%
%
\end{document}